\documentclass[10pt,conference,compsocconf]{IEEEtran}

\usepackage{cite}
\usepackage{amsmath,amssymb,amsfonts}
\usepackage{algorithmic}
\usepackage{graphicx}
\usepackage{textcomp}
\usepackage{xcolor}
\usepackage{booktabs}

\usepackage{makecell} 
\usepackage{array}
\usepackage{tabularx}

\newcommand{\arch}{\textit{\textbf{FusionCIM}}}

\begin{document}

\title{FusionCIM: Accelerating LLM Inference with Fusion-Driven Computing-in-Memory Architecture}

\author{
    \IEEEauthorblockN{
        Zihao Xuan\IEEEauthorrefmark{1}\IEEEauthorrefmark{2}, 
        Jia Chen\IEEEauthorrefmark{3}, 
        Yewen Li\IEEEauthorrefmark{1}\IEEEauthorrefmark{2}, 
        Wei Xuan\IEEEauthorrefmark{1}\IEEEauthorrefmark{2}, 
        Hegan Chen\IEEEauthorrefmark{1}, 
        Xiao Huo\IEEEauthorrefmark{1}, 
        Fengbin Tu\IEEEauthorrefmark{2}\IEEEauthorrefmark{4}
    }
    \IEEEauthorblockA{\IEEEauthorrefmark{1}AI Chip Center for Emerging Smart Systems (ACCESS), InnoHK, Hong Kong}
    \IEEEauthorblockA{\IEEEauthorrefmark{2}The Hong Kong University of Science and Technology, Hong Kong, China}
    \IEEEauthorblockA{\IEEEauthorrefmark{3}Huazhong University of Science and Technology, Wuhan, China}
    \IEEEauthorblockA{\IEEEauthorrefmark{4}Corresponding author: fengbintu@ust.hk}
}

\maketitle

\begin{abstract}

In this paper, we propose \arch, an operator-fusion-driven compute-in-memory (CIM) accelerator architecture for efficient and scalable LLM inference, with three key innovations: (1) a hybrid CIM pipeline architecture that maps $QK^T$ computation on inner-product-based CIM (IP-CIM) and $PV$ aggregation on outer-product-based CIM (OP-CIM) for efficient matrix multiplications fusion; (2) a QO-stationary dataflow that eliminates repeated KV loading in CIM and K-matrix access in buffer under transpose fusion, significantly improving data reuse on chip; and (3) a pattern-aware online-softmax mechanism that exploits distribution regularities of attention scores to reduce exponential rescaling overhead for non-linear fusion. Experimental results on LLaMA-3 model show that \arch{} achieves up to 3.86× energy saving, and 1.98× speedup compared with prior SOTA CIM-based designs with 29.4 TOPS/W energy efficiency at the system level.

\end{abstract}

\section{Introduction}

Large Language Models (LLMs) have achieved remarkable progress across various AI applications, including natural language processing (NLP), code generation, and multimodal understanding~\cite{brown2020language, wu2024next}. However, executing LLM inference efficiently on hardware remains a major limitation due to the massive computation and memory access demands during both the prefilling and autoregressive decoding phases~\cite{yu2024cambricon}. Current commercial solutions, such as GPU and TPU, rely heavily on off-chip DRAM, leading to high energy consumption and severe bandwidth bottlenecks~\cite{choquette2021nvidia, norrie2021design} .

\begin{figure*}[t]
    \centering
    \includegraphics[width=1\linewidth]{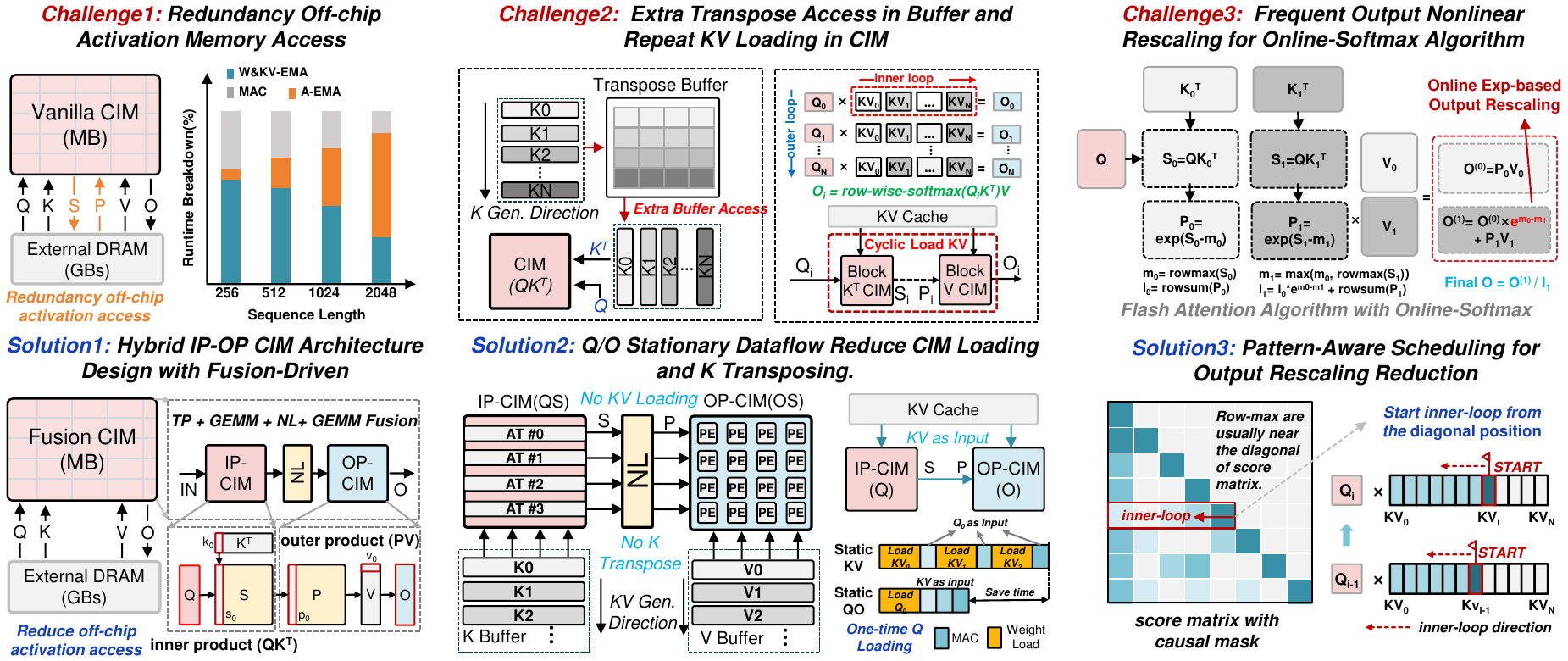}
    \caption{ Three challenges in current CIM architecture for LLM and corresponding solutions in \arch{}.}
    \label{fig:challenge}
\end{figure*}

Compute-in-Memory (CIM) architectures have emerged as a promising solution for LLM acceleration by integrating arithmetic logic (e.g., multiplier and adder) directly within or near memory arrays, greatly reducing weight data movement and improving energy efficiency \cite{chih202116, xuan2022brain}. Nevertheless, as shown in Fig.~\ref{fig:challenge}, the current CIM solutions still face several critical challenges that significantly limit their performance and energy efficiency, particularly under long-context LLM workloads.

\begin{figure}
    \centering
    \includegraphics[width=1\linewidth]{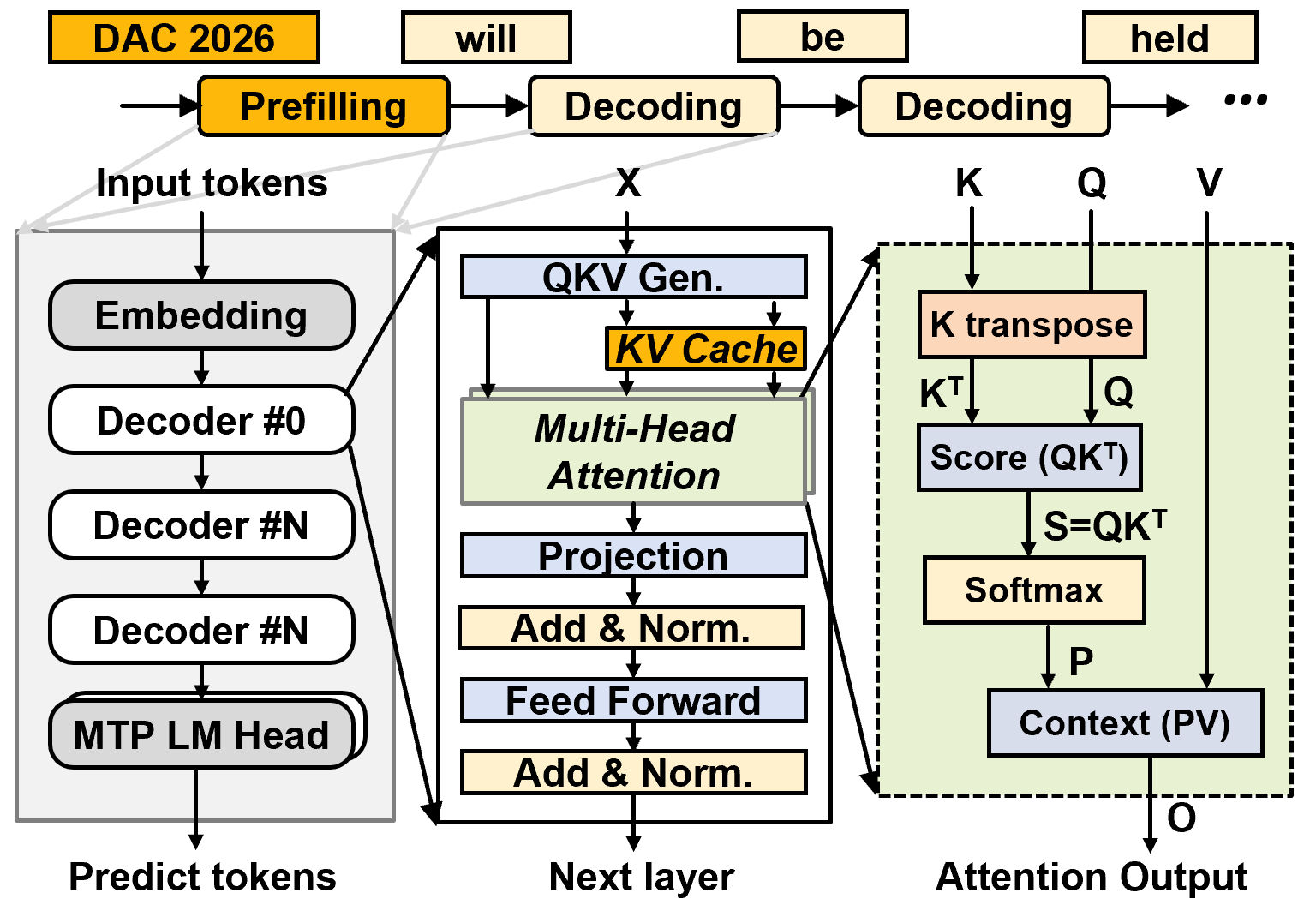}
    \caption{Model framework and inference in LLM.}
    \label{fig:LLM_algorithm}
    \vspace{-4mm}
\end{figure}

\textbf{Challenge 1: High off-chip activation access.} CIM architectures effectively reduce weight-loading overhead, but they provide limited benefit for activation data, which still requires frequent off-chip access. As the attention context length (N) grows, the overhead of activation matrices access, such as the quadratic-size score matrix (O(N$^2$)), become the primary performance bottleneck, contributing over 60\% of total latency when the N exceeds 2K. 

\textbf{Challenge 2: High on-chip KV access from tiling algorithm and transpose operations.} Due to limited on-chip memory capacity, attention is typically executed in a tiled manner \cite{leviathan2023fast}. In existing designs, the tiled KV matrices are  loaded into CIM arrays as stationary weights~\cite{tu2022trancim, tu202316, fu2023p}. However, all KV tiles must be repeatedly loaded multiple times under different Q tiles, leading to poor data reuse in CIM. Since CIM arrays suffer much higher write energy and latency than read, this repeated loading introduces substantial overhead and degrades its efficiency. Moreover, current CIM architectures require extra customized on-chip buffer or circuit to support K-matrix transposition, leading to additional on-chip memory access and area overhead~\cite{wang2025syscim, park2024tp}.

\textbf{Challenge 3: High computational overhead for online softmax algorithm.} To fuse the softmax operation with tile-wise attention computation, online-softmax is commonly adopted~\cite{dao2022flashattention}. However, it requires repeated exponential rescaling and accumulation across output partial sums, imposing heavy workload on nonlinear units and introducing extra latency.

To address these challenges, this paper presents \arch{}, a hybrid CIM-based LLM accelerator. It combines a fusion-driven architecture with optimized dataflow and pattern-aware scheduling algorithm to reduce both off-chip and on-chip memory accesses as well as nonlinear computational overhead across attention computation.

The main contributions are summarized as follows:

\begin{enumerate}

    \item \textbf{Hybrid CIM architecture with operator fusion.} A hybrid architecture integrating inner-product (IP) and outer-product (OP) CIM macros with on-chip transpose and nonlinear support, enabling GEMM fusion, transpose fusion, and nonlinear fusion to reduce redundant off-chip activation access.
    
    \item \textbf{QO-stationary dataflow.} A optimized dataflow that keeps query (Q) and attention output (O) stationary in hybrid CIM engine, maximizing data reuse and enabling in-situ K transposition without additional buffers or permutation hardware.
    
    \item \textbf{Pattern-aware online-softmax.} A softmax-friendly scheduling strategy that running $QK^T$ start form diagonal position of score to identify row maxima earlier, reducing exponential rescaling operations of attention output and alleviating nonlinear units workload.

    \item \textbf{A specific evaluation of the architecture shows} \arch{} achieves up to 3.86× energy saving, and 1.98× speedup compared with prior SOTA CIM-based designs with 29.4 TOPS/W energy efficiency and 2.03 TOPS/mm\textsuperscript{2} area efficiency at the system-level.
    
\end{enumerate}



\section{Background and Motivation}

\begin{figure*}[t]
    \centering
    \includegraphics[width=1\linewidth]{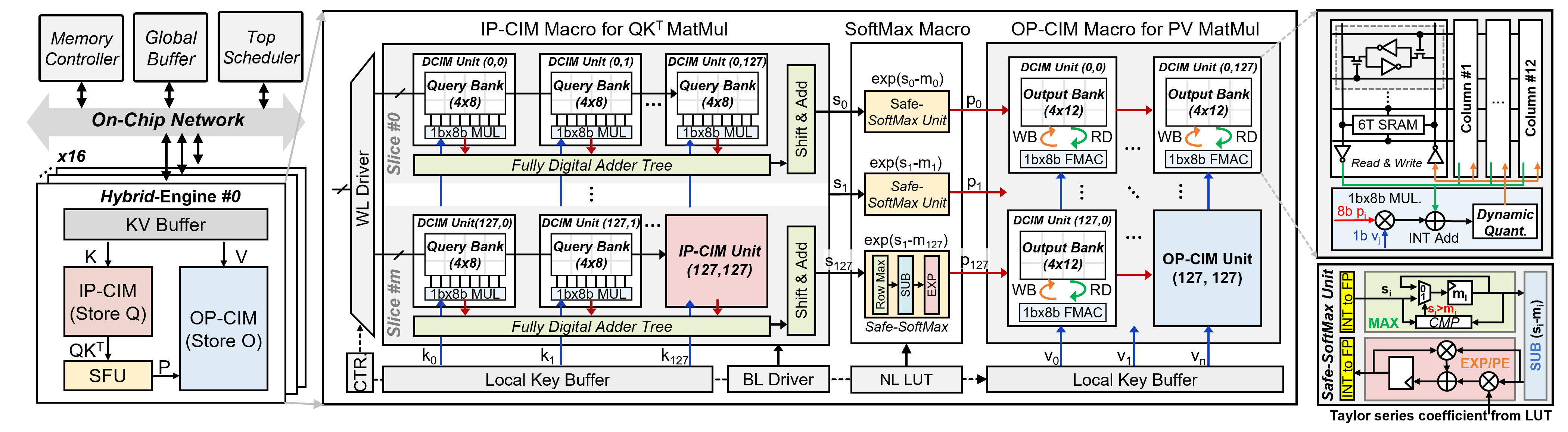}
    \caption{The overall architecture of \arch{}.}
    \label{fig:overallarch}
\end{figure*}

\subsection{LLM inference}\label{AA}
As shown in Fig.~\ref{fig:LLM_algorithm}, Large Language Models (LLMs) are typically built upon the transformer architecture, which stacks multiple decoder layers to process sequential input tokens \cite{vaswani2017attention}. Each decoder layer contains two main parts: the multi-head attention (MHA) module, which performs the $QK^T$ (score), softmax, and $PV$ (context) operations, and several fully connected layers for $QKV$ Generation, attention projection, and feed-forward network (FFN).

During inference, the model operates in two main phases: the Prefilling phase and the decoding phase. In the prefilling phase, the model processes all prompt tokens in parallel to construct the initial KV cache required for subsequent token generation. This stage is compute-intensive, involving dense matrix multiplications and attention operations, and requires large on-chip compute power. In contrast, the decoding phase generates output tokens sequentially while incrementally updating the KV cache. For each new token, the model must access the full set of weights and the history of the KV. This leads to a significant memory access problem, making the decoding phase primarily memory-bound. To alleviate this, recent algorithms have proposed parallel decoding strategies, such as speculative decoding, multi-latent attention (MLA), and multi-token prediction (MTP), allowing multiple tokens to be generated concurrently and improving decoding efficiency \cite{leviathan2023fast, liu2024deepseek, gloeckle2024better}.


\subsection{Computing-in-Memory}


Computing-in-memory (CIM) integrates computational logic within memory arrays to perform efficient vector-matrix multiplication (VMM) with fixed weights. Depending on the implementation, CIM can be classified into analog and digital designs. Digital CIM (DCIM), benefiting from high precision and process scaling, has become the dominant architecture in modern CIM designs. In a typical DCIM, each memory cell is equipped with a digital multiplier to multiply input data with pre-loaded static weights, while each output channel contains an adder tree for parallel accumulation~\cite{chen2023autodcim}. During computation, the input vector is broadcast along the rows, and the multiplication results are accumulated along the output direction, enabling a VMM in a single cycle. By tightly coupling computation with storage, CIM architectures substantially reduce data-movement overhead, thereby enhancing energy efficiency and throughput while mitigating the performance limitations of the traditional memory wall.


\subsection{Motivation}

Previous CIM designs primarily focus on macro-level optimizations of computational circuit and weight reuse. However, in the context of large language models (LLMs), on-chip CIM is often unable to cache all weights and KV pairs, making memory access the primary system bottleneck and limiting overall CIM efficiency. Recent CIM accelerators, such as TranCIM~\cite{tu2022trancim}, attempt to address this issue by adopting pipeline CIM architectures to reduce activation memory access. Nevertheless, these approaches typically map KV matrices to CIM as static weights while treating Q matrices as inputs for attention computation. This strategy introduces several challenges: KV matrices must be repeatedly read and written within CIM; on-chip K-matrix transposition produces additional memory access or circuit overhead. In addition, the nonlinear softmax operation introduces extra latency in online processing due to repeated exponential rescaling.

These challenges motivate our design a hybrid accelerator architecture that combines IP-CIM and OP-CIM to implement pipeline attention computation with a QO-stationary dataflow and pattern-aware scheduling strategy. High-efficiency IP-CIM macros handle Q-Stationary $QK^T$-MM, while flexible OP-CIM macros perform O-Stationary $PV$-MM. This hybrid design not only maximizes data reuse but also avoids repeated KV loading, and on-chip K transposition, thereby improving overall computational efficiency. Moreover, exploiting the spatial regularity of attention score distributions, a pattern-aware scheduling mechanism for online-softmax reduces normalization overhead and enhances throughput. Overall, this hybrid CIM design achieves scalable, memory-efficient, and high-performance inference for large-scale LLMs.

\section{\arch{} Architecture}


\subsection{Overall Architecture}

The proposed \arch{} architecture adopts a hierarchical, memory-centric organization to efficiently support large-scale attention computation in LLM inference. As shown in Fig.~\ref{fig:overallarch}, the architecture consists of a Global Buffer, a Top Scheduler, a Memory Controller, and multiple Hybrid Engines (HE\#0$\sim$HE\#15). The Global Buffer temporarily caches Key (K) and Value (V) matrices to reduce off-chip DRAM access. The Top Scheduler coordinates workload distribution across hybrid engines, enabling flexible pattern-aware scheduling and dynamic load balancing across attention heads or data tiles. The Memory Controller interfaces with external DRAM and performs asynchronous data prefetching to sustain high throughput. Each HE executes pipelined, tile-wise attention with a Query (Q) and Output (O) stationary dataflow. This modular organization enables scalable parallelism and high utilization, making \arch{} well-suited for both the prefill phase and the parallel decode phase.

\subsection{Hybrid Engine Design}

Each Hybrid Engine integrates three core modules: a IP-CIM Macro that performs \emph{inner-product} computation for $QK^T$-MM (score), a OP-CIM Macro that performs \emph{outer-product} computation for $PV$-MM (context), and a lightweight SoftMax core that executes \emph{online-softmax} for normalization. These modules are tightly connected to enable pipelined attention computation within each engine.

\subsubsection{IP-CIM Macro for $QK^T$ Computation}

The basic building unit of the IP-CIM macro consists of a 4$\times$8 SRAM bank and a 1-bit$\times$8-bit multiplier, and these units are replicated and organized into a 2D array to form the complete CIM structure. The IP-CIM macro performs tiled score computations with a $Q$-stationary dataflow, keeping the pre-loaded $Q$ matrix in local SRAM banks while streaming $K$ vectors from the K buffer into the CIM array. During computation, one wordline(WL) in the SRAM bank of basic unit is activated to read out an 8-bit $Q$, while the corresponding 1-bit $K$ value is broadcast bit-serially along the column direction. Each multiplier performs 1b$\times$8b multiplication, and the outputs within the same row are accumulated by a full digital adder-tree circuit. To minimize data movement, the CIM periphery integrates shift-and-add logic, enabling in-array partial summation. The accumulated $QK^T$ scores are forwarded directly to the SoftMax unit without intermediate buffering.

\subsubsection{OP-CIM Macro for $PV$ Computation}

The OP-CIM macro performs tiled $PV$ computations using an $O$-stationary dataflow. Its basic building unit consists of a 4$\times$16 SRAM bank and a fused 1b$\times$8b MAC (FMAC). During execution, the SoftMax output vector ($P$) is broadcast along the row direction to the FMAC units, while the $V$ vectors are streamed bit-serially along the column direction. Each FMAC computes the element-wise product $P_i \cdot V_j$ and accumulates it with the previous partial sum, yielding
\begin{equation}
O_{ij}[t] = P_i \times V_j + O_{ij}[t-1].
\end{equation}
The accumulated results are written back into the local SRAM bank with minimized output Psum movement. To further reduce the output bit-width required for storing attention results, the FMAC adopts a dynamic partial-sum quantization scheme similar to the method proposed in~\cite{hu202528nm}, enabling compact in-situ accumulation with minimal precision loss. 

\subsubsection{Lightweight SoftMax Macro}

A lightweight SoftMax macro is placed between the two CIM macros to perform online normalization of attention scores. Each output channel of the IP-CIM is paired with an independent SoftMax unit, enabling fully parallel processing. Each unit contains a maximum tracker, a subtractor, and an exponential computation unit. The maximum tracker maintains the running row-wise maximum of the score. 
To prevent numerical overflow in the exponentiation step, each score is adjusted by subtracting the row maximum. The exponential unit employs a Taylor expansion approach, implemented via a lightweight PE whose coefficients are stored in a local LUT, enabling energy-efficient realization.

\subsection{QO-Stationary Dataflow and Attention Computing Pipeline}

\begin{figure}
    \centering
    \includegraphics[width=\linewidth]{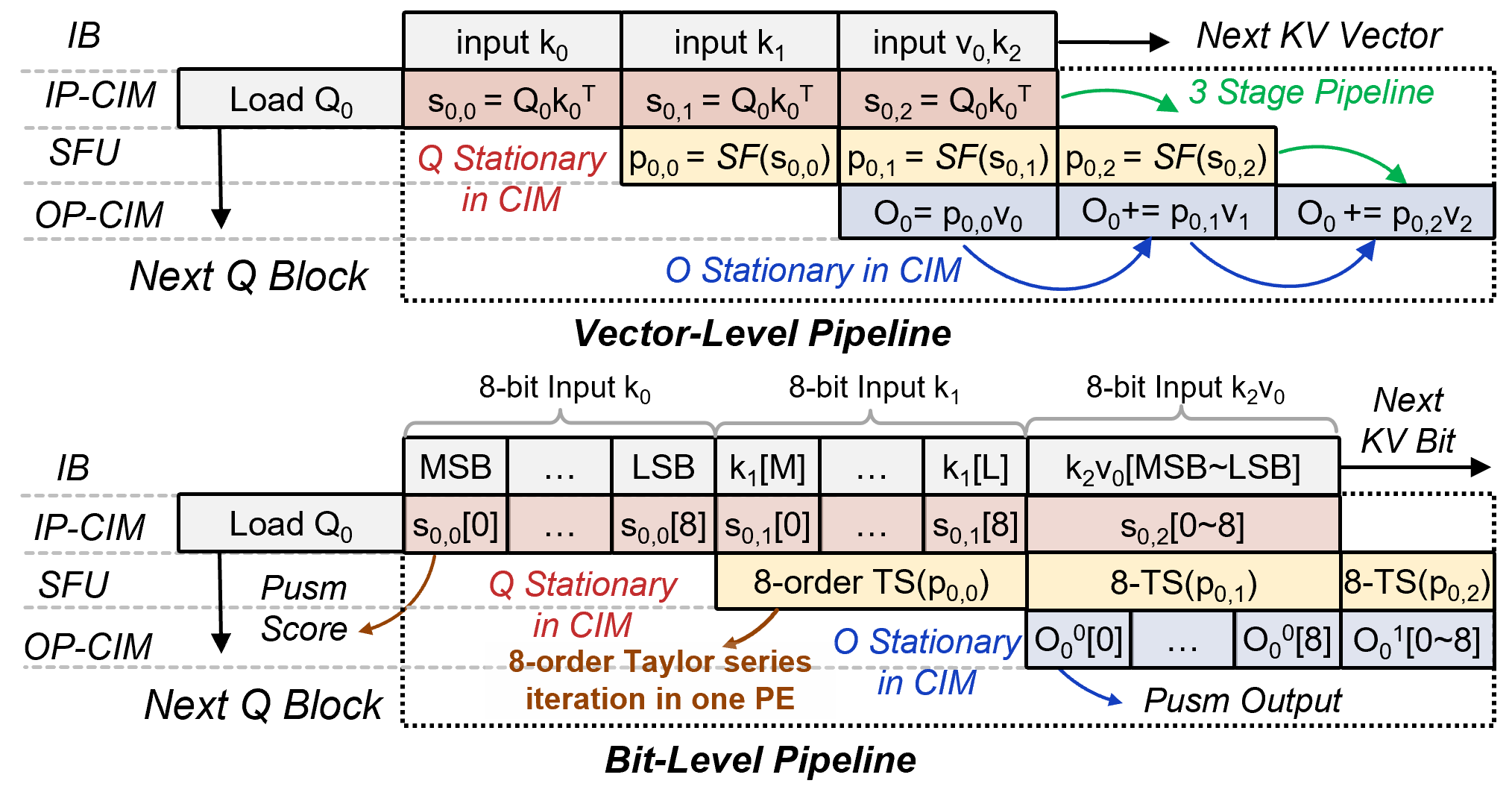}
    \caption{Two-level pipelined attention execution with single Q loading.}
    \label{fig:pipeline}
\end{figure}

During attention execution, each HE adopts a QO-stationary dataflow that maximizes data reuse and minimizes off-chip memory access. The $Q$ matrix is loaded once into the IP-CIM array and remains stationary throughout the processing of all $K$ and $V$ vectors. The IP-CIM performs the $QK^T$ computation, while the OP-CIM carries out the $PV$ aggregation. Partial results are accumulated directly within the local SRAM banks of the OP-CIM, avoiding unnecessary data transfers between pipeline stages.

As illustrated in Fig.~\ref{fig:pipeline}, the overall attention computation is organized into two coordinated pipeline levels: (1) At the \textit{vector level pipeline}, the $K$ and $V$ are broadcast vector-by-vector to HE, forming a three-stage pipeline of $QK^T$, SoftMax, and $PV$. This enables continuous vector-level throughput, as a new KV vector can enter the pipeline once the previous one advances to the next stage.
(2)Within each vector, the \textit{bit-level pipeline} further exploits the bit-serial representation of $K$ and $V$. Each 8-bit element is processed sequentially across eight cycles, forming a fine-grained bit-level pipeline inside both the IP-CIM (for $QK^T$) and the OP-CIM (for $PV$). To maintain cycle alignment, the SoftMax exponential operation is implemented using an 8-step Taylor-series iteration, allowing its latency to match the bit-serial computation of two types of CIM. By combining continuous vector-level streaming with tightly aligned bit-level serial computation, the HE sustains fully pipelined attention execution with high utilization and minimal data movement.

\subsection{Pattern-Aware Scheduling}
\begin{figure}
    \centering
    \includegraphics[width=1\linewidth]{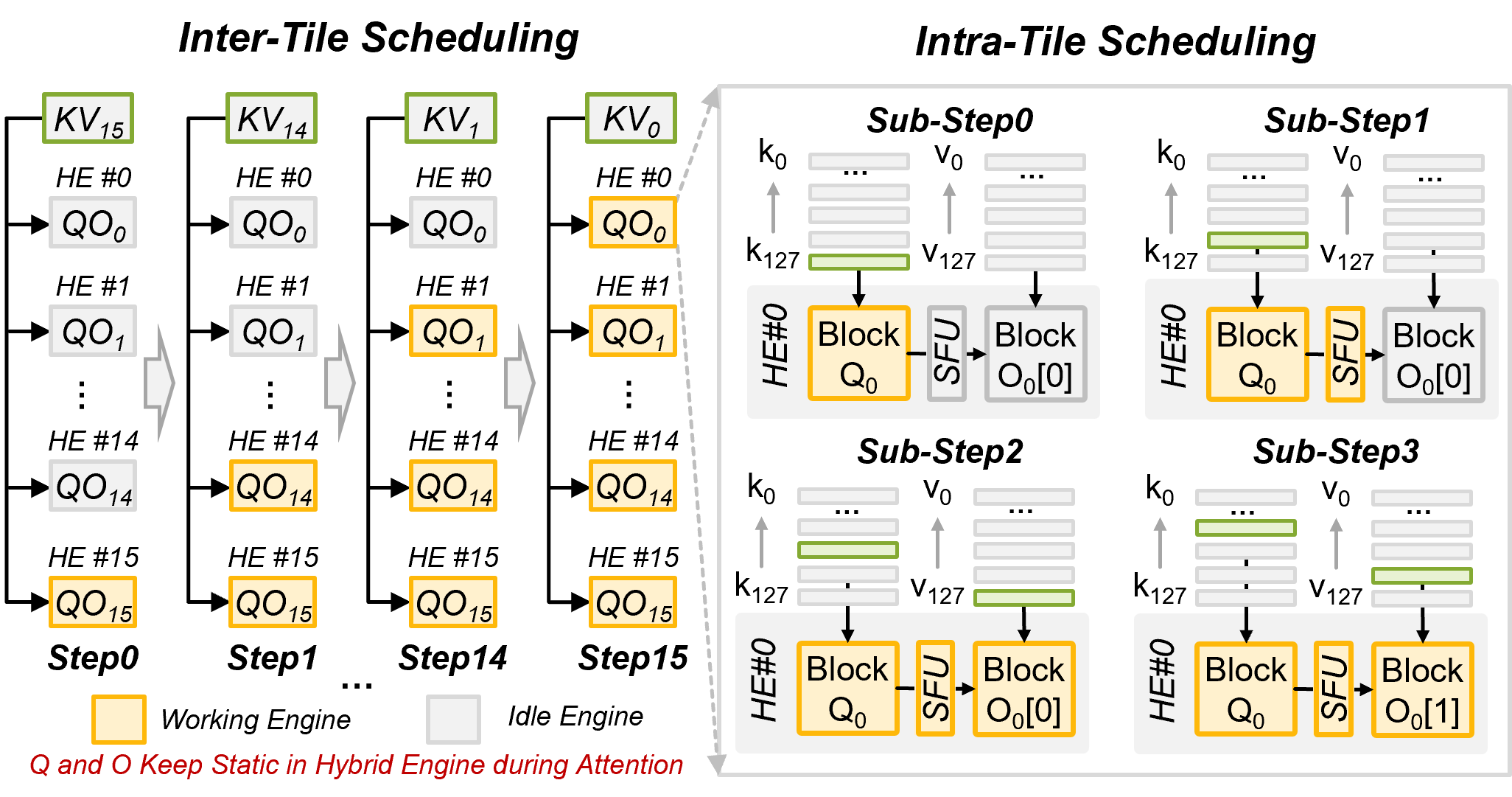}
    \caption{Inter- and intra-tile pattern-aware scheduling to reduce online-softmax rescaling.}
    \label{fig:pattern_scheduling}
\end{figure}

In the proposed \arch{} architecture, safe-SoftMax is executed at every inner-loop step of attention computation. To ensure numerical stability and match the standard safe-SoftMax results, the row-wise maximum of the score matrix must be updated at each step. This repeated update introduces frequent exponential rescaling of attention output, increasing the workload of the non-linear unit and limiting computational throughput. 

A distribution analysis of attention scores shows that row maxima are highly likely to appear near the diagonal. \arch{} exploits this property through a Pattern-Aware Scheduling strategy that initiates inner-loop computation from the diagonal, allowing early identification of row maxima and reducing unnecessary output rescaling operations.

As illustrated in Fig.~\ref{fig:pattern_scheduling}, the scheduling operates at two levels: (1)~\textit{At the inter-tile scheduling}, the complete KV matrix is divided into $N$ tiles (KV$_0$ to KV$_{N-1}$), each containing 128 KV vectors. These tiles are broadcast to multiple HEs in a staggered manner, where HE$_N$ receives KV$_N$ at step~0, KV$_{N-1}$ is broadcast to HE$_N$ and HE${_N-1}$ at step~1, until all tiles are distributed. (2)~\textit{ At the intra-tile scheduling}, each HE applies a local pattern-aware scheduler that processes KV vectors in reverse index order from $k_{127}$ down to $k_0$. 

This diagonal-first traversal captures row-wise maxima earlier, reducing redundant maximum updates during online-softmax. On LLaMA-3, this approach eliminates up to 61.4\% of output rescaling without compromising numerical accuracy (see Fig.~\ref{fig:pattern}), resulting in a measurable improving the overall throughput of \arch{}.

\section{EVALUATION}

\subsection{Experiment Setup}

\subsubsection{Architecture Configure}

\begin{table}[!t]
\centering
\caption{Architecture configuration of the proposed {\arch} accelerator.}
\label{tab:arch_config}
\resizebox{0.48\textwidth}{!}{%
\renewcommand{\arraystretch}{1.2} 
\setlength{\tabcolsep}{3pt}       
\begin{tabular}{cccc}
\toprule
\textbf{Component} & 
\makecell[c]{ \textbf{Size/Parameters} \\ \textbf{@400MHz}} & 
\textbf{Power (mW)} & \textbf{Area (mm$^2$)} \\
\midrule
System & \makecell[c]{16 Hybrid Engines (HEs);\\1 MB Global Buffer} & 2100 & 26.2 \\
\addlinespace[3pt]
Hybrid Engine & \makecell[c]{1 IP-CIM; 1 OP-CIM;\\1 Softmax Engine} & 113 & 1.24 \\
\addlinespace[3pt]
\makecell[c]{IP-CIM Macro \\ (INT8)} & \makecell[c]{128$\times$128 MACs;\\64 KB SRAM; 1.64 TOPS} & 42.0 & 0.46 \\
\addlinespace[3pt]
\makecell[c]{OP-CIM Macro \\ (INT8)} & \makecell[c]{128$\times$128 MACs;\\128 KB SRAM; 1.64 TOPS} & 53.6 & 0.68 \\
\addlinespace[3pt]
SFU Macro (FP16) & \makecell[c]{128 Units;\\128B LUT; 0.2 TOPS} & 13.4 & 0.07 \\
\bottomrule
\end{tabular}%
}
\end{table}

The proposed \arch{} accelerator is composed of 16 hybrid engines (HEs), which are interconnected through an on-chip Network-on-Chip (NoC) and a 1 MB shared global buffer. Each HE serves as an independent computational unit integrating a IP-CIM macro, a OP-CIM macro, and a softmax macro, enabling pipelined execution of tile-wise attention computation. Among them, The IP-CIM macro adopts a 128$\times$128 array structure, providing 64 KB on-core memory and achieving 1.64 TOPS throughput at 400 MHz and INT8 precision, with 42 mW power and 0.46 mm\textsuperscript{2} area.The OP-CIM macro features a similar array with 128 KB memory, delivering the same 1.64 TOPS performance while consuming 53.6 mW and occupying 0.68 mm\textsuperscript{2}. The Softmax macro includes 128 parallel FP16 processing elements (PEs), offering a throughput equivalent to 0.2 TOPS with 13.1 mW power and 0.07 mm\textsuperscript{2} area. Together, these macros form a balanced hybrid engine that supports pipelined attention computation with QO-stationary dataflow and pattern-aware online-softmax optimization for efficient LLM inference. Table.~\ref{tab:arch_config} summarizes the specific power and area of different components under a given configuration. 

\subsubsection{Architecture Simulation}

The hardware area and power characteristics are obtained through a combination of mature CAD tools and silicon measurements. Specifically, CIM module parameters are derived from open-source CIM modeling tool \cite{peng2019dnn+}; SRAM parameters are extracted using Cacti 7.0 \cite{balasubramonian2017cacti}; the compute units in softmax macro, as well as the Top Controller, are synthesized using Synopsys Design Compiler, with their energy characteristics evaluated through PrimeTime PX. All modules are analyzed under TSMC 28 nm CMOS technology, where the post-synthesis critical path delay reaches 2.2 ns, supporting a 400 MHz operating frequency for the \arch{} accelerator. 

\subsubsection{Hardware Baselines}

To demonstrate the advantages of the proposed \arch{} architecture, we implement and compare against two baseline architectures with equivalent computational capability. \textit{Baseline 1: DCIM-based Architecture.} Each HE just contains a conventional digital CIM macro with 128$\times$256 size to match the overall computational throughput of our design. \textit{Baseline 2: TransCIM-like Architecture.} Each HE contains a pair of vanilla CIM cores, each with 128$\times$128 MAC units, performing pipeline attention computation with static KV mapping similar to TranCIM~\cite{tu2022trancim}. The transpose operations in Baseline~1 is implemented via additional SRAM buffer, and the softmax modules in both baselines are kept identical to that in \arch{} for fair comparison.

\begin{figure}[!t]
    \centering
    \includegraphics[width=\linewidth]{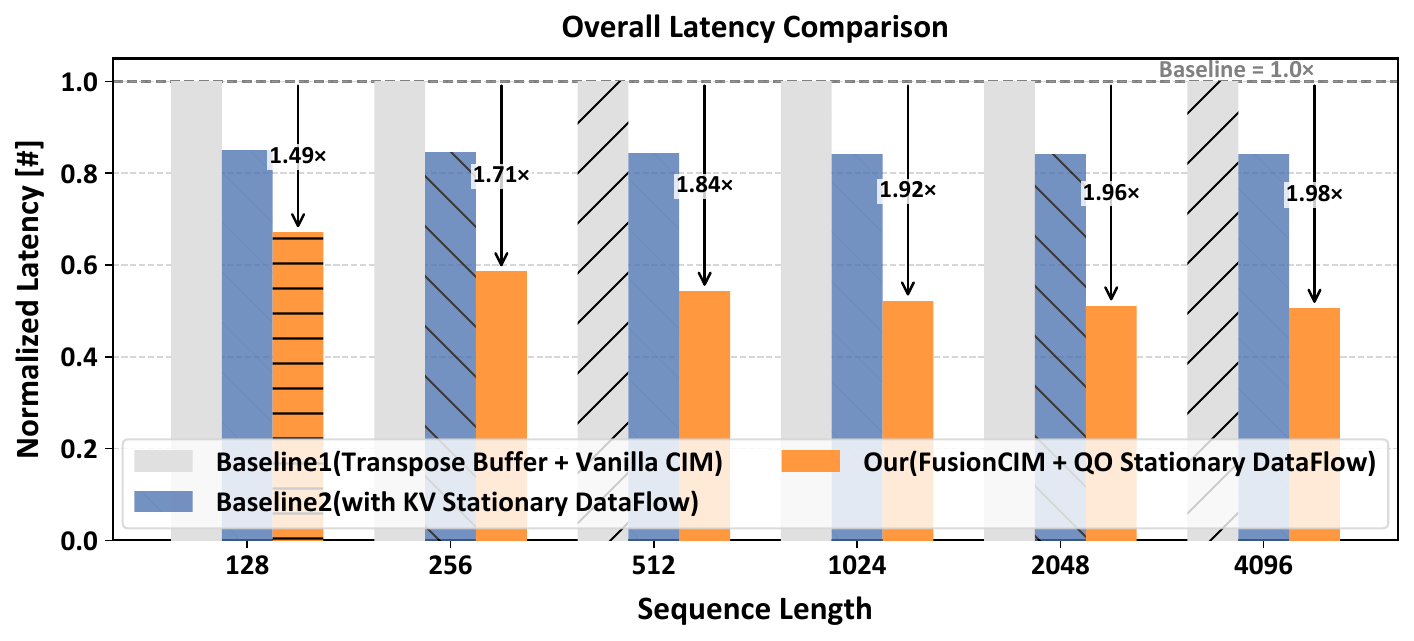}
    \caption{Normalized latency comparison between FusionCIM and two Baselines.}
    \label{fig:speedup}
\end{figure}

\begin{figure}[!t]
    \centering
    \includegraphics[width=\linewidth]{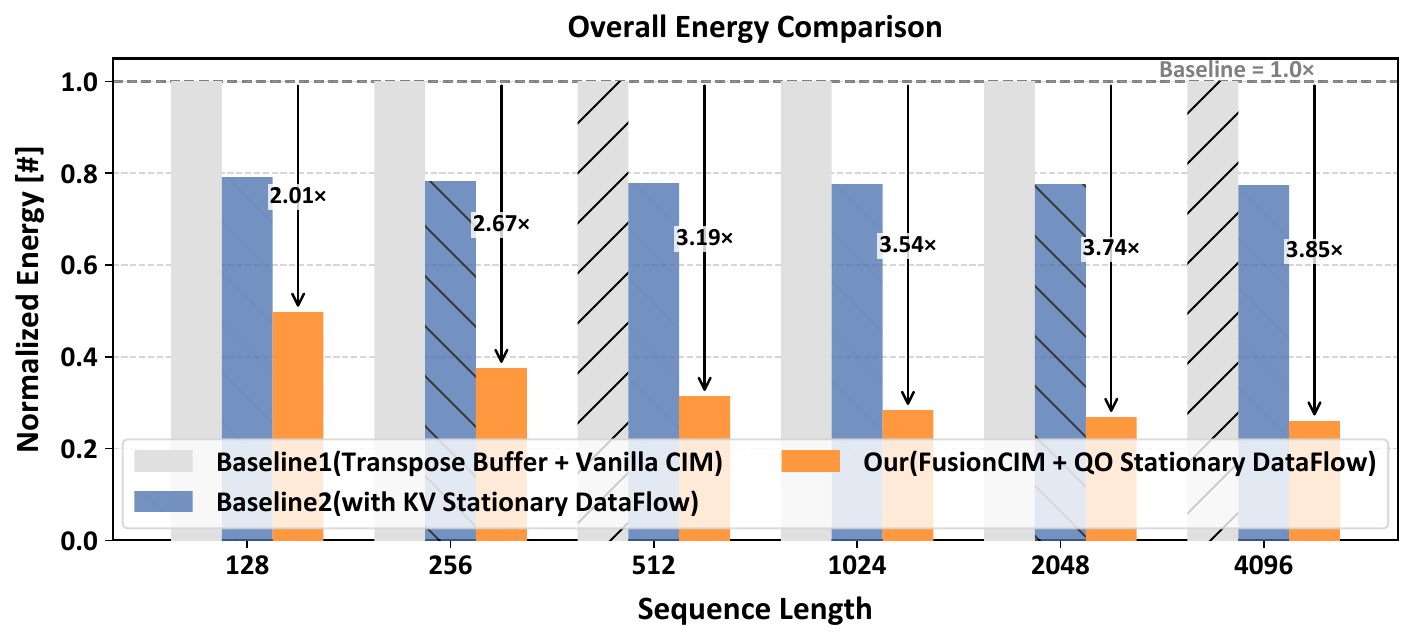}
    \caption{Normalized energy consumption comparison between FusionCIM and two Baselines.}
    \label{fig:energysaving}
    \vspace{-2mm}
\end{figure}

\subsubsection{LLM Workloads}

We employ LLaMA3-8B model as the primary workload \cite{dubey2024llama}. The model is configured with a maximum context length of 8K, a hidden dimension of 4096, and 32 attention heads using grouped-query attention (GQA). Given that other LLMs (e.g., GPT-3) share similar attention structures and head dimensions, their computational and performance characteristics can be scaled and inferred from the LLaMA3 benchmark, ensuring representative evaluation across diverse LLM architectures.

\subsection{Experiment Results}

\subsubsection{Evaluation of Acceleration Ratio }

Fig.~\ref{fig:speedup} compares the normalized latency of our design with two baselines across different sequence lengths (Baseline~1 = 1.0). The hybrid pipeline \arch{} architecture achieves significant speedup over Baseline~1, reaching up to 1.98$\times$ at a sequence length of 4K. This improvement stems from the deep operator fusion enabled by the heterogeneous pipeline, which effectively reduces latency caused by off-chip memory accesses. Compared to Baseline~2, the proposed QO-stationary dataflow further reduces latency by 21--40\% by minimizing on-chip CIM read/write operations and improving array utilization, thereby accelerating computation. Overall, these results demonstrate that the proposed optimizations substantially enhance execution efficiency and throughput, particularly for long-sequence inference.

\begin{figure}[htbp]
    \centering
    \includegraphics[width=\linewidth]{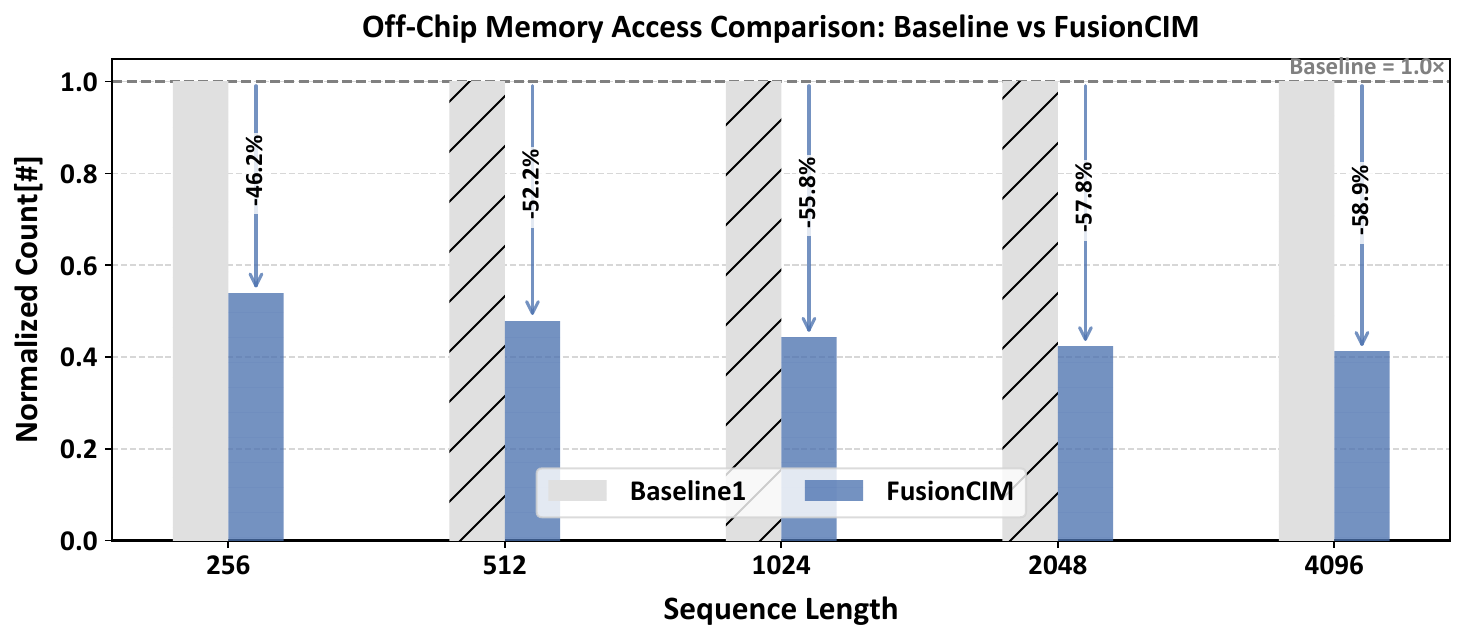}
    \caption{Normalized off-chip memory access comparison between Baseline~1 and FusionCIM.}
    \label{fig:off_chip_memory_access}
\end{figure}

\begin{figure}[htbp]
    \centering
    \includegraphics[width=1\linewidth]{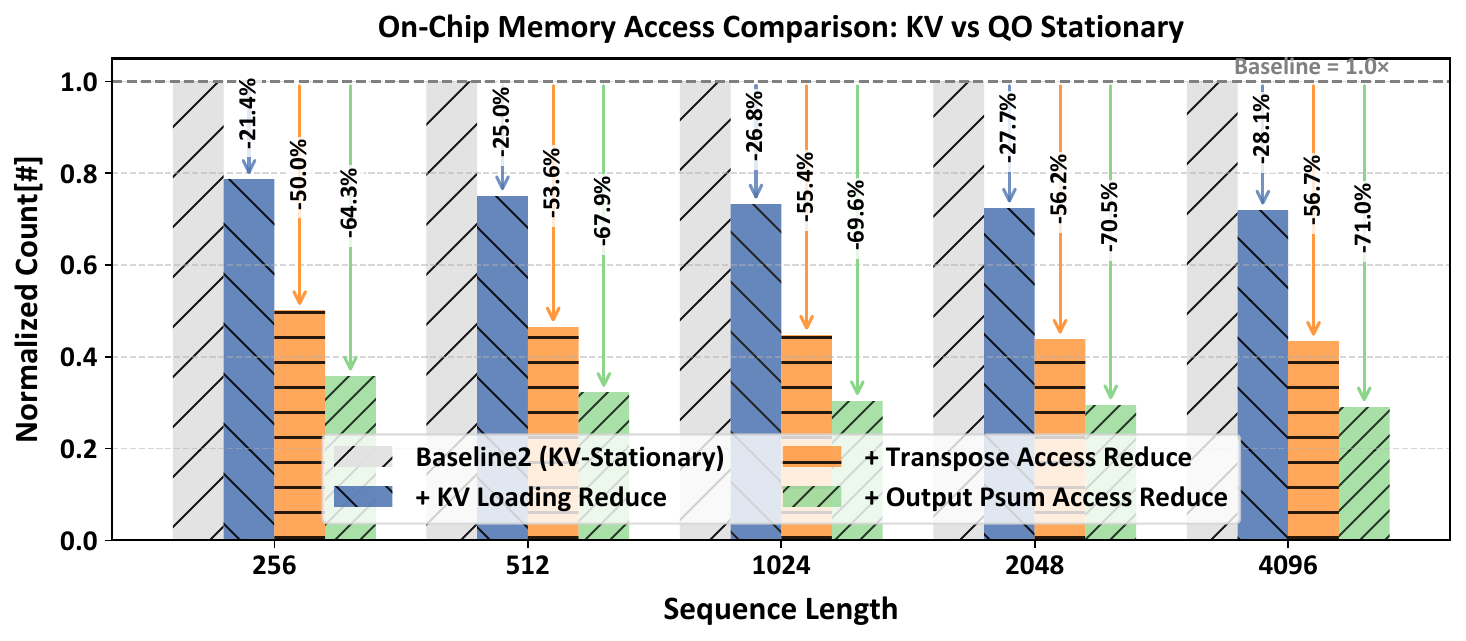}
    \caption{Normalized on-chip memory access comparison between Baseline~2 and FusionCIM, and ablation studies.}
    \label{fig:onchip_memory_access}
\end{figure}

\subsubsection{Evaluation of Energy Saving}

Fig.~\ref{fig:energysaving} shows the energy consumption normalized to Baseline~1 (1.0$\times$). Compared with two baselines, our method achieves the lowest energy usage across various sequence lengths. Moreover, the energy advantage becomes increasingly pronounced as the sequence length grows. The proposed \arch{} delivers progressively higher energy savings, ultimately reducing energy consumption by up to 3.85$\times$ at longer sequences(4K).
This trend demonstrates the high efficiency of our approach for compute-intensive workloads, making it well suited for energy-constrained edge applications.

\subsubsection{Memory Access Analysis}
To evaluate the effectiveness of our proposed architecture in reducing data movement, we analyze both off-chip and on-chip memory access behaviors. As shown in  Fig.~\ref{fig:off_chip_memory_access}, the proposed \arch{}  substantially decreases off-chip memory accesses compared to the Baseline~1, achieving a 46.2\%–58.9\% reduction as the sequence length increases from 256 to 4096. Fig.~\ref{fig:onchip_memory_access} further compares the on-chip memory access between KV-stationary (baseline~2) and QO-stationary (Our) schemes. The QO-stationary dataflow reduces on-chip memory traffic by optimizing KV loading, transpose accesses, and output partial-sum movement. To assess the contribution of each optimization, we conduct ablation studies. The results shown in Fig.~\ref{fig:onchip_memory_access} indicate that QO effectively decreases data access in all three stages, leading to an overall 64.3\%-71.0\% reduction in on-chip memory access across different sequence lengths. 

\begin{figure}[htbp]
    \centering
    \includegraphics[width=1\linewidth]{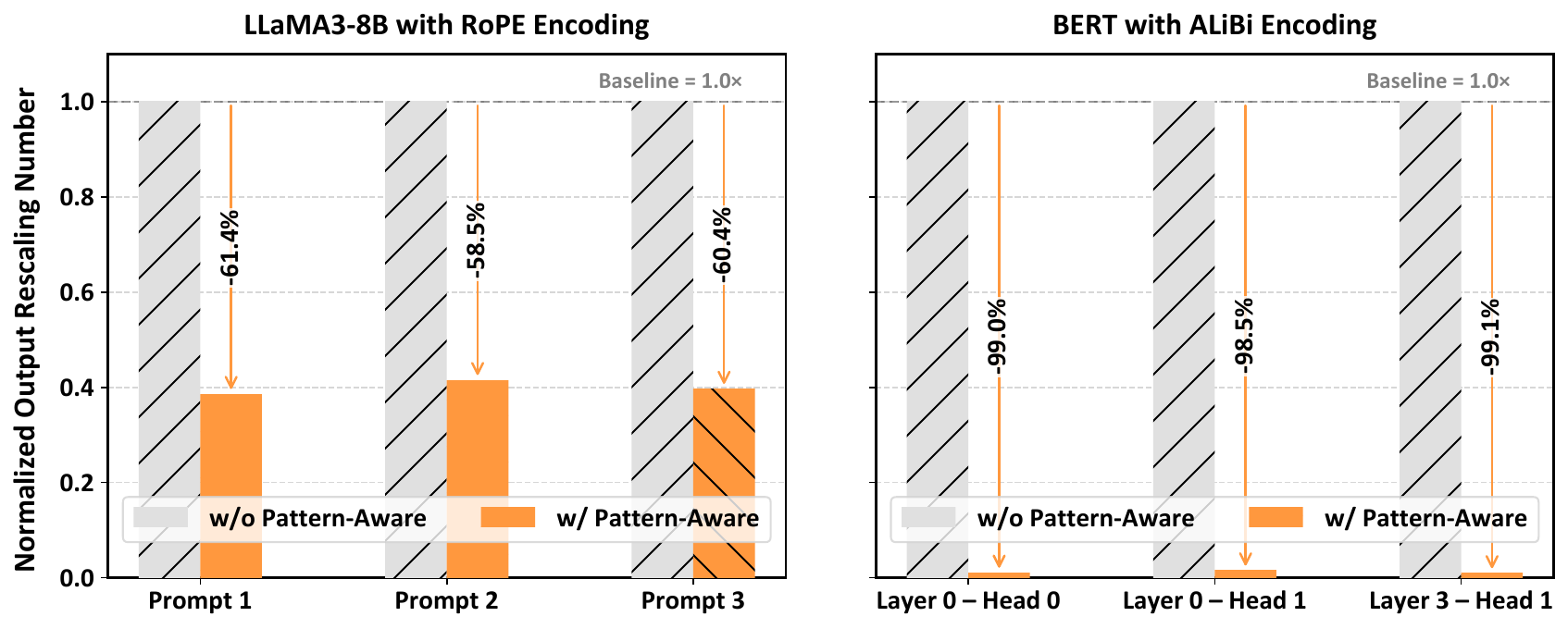}
    \caption{Normalized output rescaling number comparison under different models, prompts, layers, and heads.}
    \label{fig:pattern}
\end{figure}

\subsubsection{Output Rescaling Analysis}

Fig.~\ref{fig:pattern} evaluates the effectiveness of our pattern-aware scheduling across different models, prompts, layers, and attention heads. On LLaMA3-8B with rotary positional encoding (RoPE), our method reduces the number of output rescaling by 58.5\%--61.4\% across various prompts. Hardware-software co-design can further lower rescaling overhead, as recent techniques like ALiBi encoding are well-suited for our approach. For Bert with ALiBi encoding, the reduction is even more significant, achieving 98.5\%--99.1\% fewer rescalings. By minimizing rescaling number, we greatly reduce the computational cost of non-linear operations, enabling faster and more energy-efficient inference with negligible impact on model accuracy.

\subsubsection{Computational Density and Energy Efficiency Comparison}

\begin{table}[h]
    \centering
    \caption{Comparison to the State-of-the-Art CIM-based Accelerators.}
    \label{tab:eeperf}
    \setlength{\tabcolsep}{5pt}
    \small
    \begin{tabularx}{\columnwidth}{|>{\raggedright\arraybackslash}p{0.25\columnwidth}
                                     |>{\centering\arraybackslash}X
                                     |>{\centering\arraybackslash}X
                                     |>{\centering\arraybackslash}X|}
        \hline
         & P3ViT \cite{fu2023p} & TranCIM \cite{tu2022trancim} & This Work \\ \hline
        Technology (nm) & 28 nm & 28 nm & 28 nm \\ \hline
        Dataflow Mode  & KV-Static & KV-Static & \textbf{QO-Static} \\ \hline
        Transpose   & No  & Yes & \textbf{Yes (native) }\\ \hline
        Softmax Support      & No  & No & \textbf{Yes} \\ \hline
        Frequency & 200 MHz & 240 MHz & 400 MHz \\ \hline
        Precision & INT4/8/16 & INT8/16 & INT4/8 \\ \hline
        \makecell[l]{Area Efficiency \\ (TOPS/mm$^2$)} 
            & \makecell[c]{0.8 \\ (INT8)} 
            & \makecell[c]{0.22 \\ (INT8)} 
            & \makecell[c]{\textbf{2.03} \\ (INT8)} \\ \hline

        \makecell[l]{Energy Efficiency \\ (TOPS/W)} 
            & \makecell[c]{23.2(Macro) \\ (INT8)} 
            & \makecell[c]{12.5(System) \\ (INT8)} 
            & \makecell[c]{\textbf{29.4(Sys.)} \\ (INT8)} \\ \hline
    \end{tabularx}
\end{table}

Table.~\ref{tab:eeperf} presents the comparison with the state-of-the-art DCIM-based accelerators for LLM acceleration. The proposed hybrid pipeline CIM accelerator improves energy efficiency by 1.26$\times$ and 2.35$\times$, and enhances area efficiency by 2.5$\times$ and 9.2$\times$, compared to the recent P3ViT \cite{fu2023p} and TranCIM \cite{tu2022trancim} , respectively.


\section{Conclusion}

In this paper, we present \arch{}, an operator-fusion-driven architecture that integrates a hybrid CIM design, optimized dataflow, and pattern-aware scheduling to maximize data reuse while minimizing memory access and nonlinear computation overhead. Our architecture achieves 3.85$\times$ energy savings and 1.98$\times$ speedup, with 29.4 TOPS/W energy efficiency and 2.03 TOPS/mm\textsuperscript{2} area efficiency at the system level. These results demonstrate that \arch{} provides a compelling alternative solution for future high-performance LLM accelerators.

\section{Acknowledgment}

This research was conducted by ACCESS – AI Chip Center for Emerging Smart Systems, supported by the InnoHK initiative of the Innovation and Technology Commission of the Hong Kong Special Administrative Region Government.

\bibliographystyle{IEEEtran}
\bibliography{main}

@article{brown2020language,
  title={Language models are few-shot learners},
  author={Brown, Tom and Mann, Benjamin and Ryder, Nick and Subbiah, Melanie and Kaplan, Jared D and Dhariwal, Prafulla and Neelakantan, Arvind and Shyam, Pranav and Sastry, Girish and Askell, Amanda and others},
  journal={Advances in neural information processing systems},
  volume={33},
  pages={1877--1901},
  year={2020}
}

@article{tu2022trancim,
  title={TranCIM: Full-digital bitline-transpose CIM-based sparse transformer accelerator with pipeline/parallel reconfigurable modes},
  author={Tu, Fengbin and Wu, Zihan and Wang, Yiqi and Liang, Ling and Liu, Liu and Ding, Yufei and Liu, Leibo and Wei, Shaojun and Xie, Yuan and Yin, Shouyi},
  journal={IEEE Journal of Solid-State Circuits},
  volume={58},
  number={6},
  pages={1798--1809},
  year={2022},
  publisher={IEEE}
}

@article{fu2023p,
  title={P 3 ViT: A CIM-based high-utilization architecture with dynamic pruning and two-way ping-pong macro for vision transformer},
  author={Fu, Xiangqu and Ren, Qirui and Wu, Hao and Xiang, Feibin and Luo, Qing and Yue, Jinshan and Chen, Yong and Zhang, Feng},
  journal={IEEE Transactions on Circuits and Systems I: Regular Papers},
  volume={70},
  number={12},
  pages={4938--4948},
  year={2023},
  publisher={IEEE}
}

@inproceedings{hu202528nm,
  title={A 28nm 20.9-137.2 TOPS/W Output-Stationary SRAM Compute-in-Memory Macro Featuring Dynamic Look-ahead Zero Weight Skipping and Runtime Partial Sum Quantization},
  author={Hu, Xiaofeng and Mun, HanGyeol and Meng, Jian and Liao, Yuan and Sridharan, Amitesh and Seo, Jae-sun},
  booktitle={2025 IEEE Custom Integrated Circuits Conference (CICC)},
  pages={1--3},
  year={2025},
  organization={IEEE}
}

@article{balasubramonian2017cacti,
  title={CACTI 7: New tools for interconnect exploration in innovative off-chip memories},
  author={Balasubramonian, Rajeev and Kahng, Andrew B and Muralimanohar, Naveen and Shafiee, Ali and Srinivas, Vaishnav},
  journal={ACM Transactions on Architecture and Code Optimization (TACO)},
  volume={14},
  number={2},
  pages={1--25},
  year={2017},
  publisher={ACM New York, NY, USA}
}

@inproceedings{yu2024cambricon,
  title={Cambricon-llm: A chiplet-based hybrid architecture for on-device inference of 70b llm},
  author={Yu, Zhongkai and Liang, Shengwen and Ma, Tianyun and Cai, Yunke and Nan, Ziyuan and Huang, Di and Song, Xinkai and Hao, Yifan and Zhang, Jie and Zhi, Tian and others},
  booktitle={2024 57th IEEE/ACM International Symposium on Microarchitecture (MICRO)},
  pages={1474--1488},
  year={2024},
  organization={IEEE}
}

@inproceedings{chih202116,
  title={16.4 An 89TOPS/W and 16.3 TOPS/mm 2 all-digital SRAM-based full-precision compute-in memory macro in 22nm for machine-learning edge applications},
  author={Chih, Yu-Der and Lee, Po-Hao and Fujiwara, Hidehiro and Shih, Yi-Chun and Lee, Chia-Fu and Naous, Rawan and Chen, Yu-Lin and Lo, Chieh-Pu and Lu, Cheng-Han and Mori, Haruki and others},
  booktitle={2021 IEEE International Solid-State Circuits Conference (ISSCC)},
  volume={64},
  pages={252--254},
  year={2021},
  organization={IEEE}
}

@article{choquette2021nvidia,
  title={Nvidia a100 tensor core gpu: Performance and innovation},
  author={Choquette, Jack and Gandhi, Wishwesh and Giroux, Olivier and Stam, Nick and Krashinsky, Ronny},
  journal={IEEE Micro},
  volume={41},
  number={2},
  pages={29--35},
  year={2021},
  publisher={IEEE}
}

@article{xuan2022brain,
  title={A brain-inspired ADC-free SRAM-based in-memory computing macro with high-precision MAC for AI application},
  author={Xuan, Zihao and Liu, Chang and Zhang, Yue and Li, Yuan and Kang, Yi},
  journal={IEEE Transactions on Circuits and Systems II: Express Briefs},
  volume={70},
  number={4},
  pages={1276--1280},
  year={2022},
  publisher={IEEE}
}

@inproceedings{tu202316,
  title={16.4 TensorCIM: A 28nm 3.7 nJ/gather and 8.3 TFLOPS/W FP32 digital-CIM tensor processor for MCM-CIM-based beyond-NN acceleration},
  author={Tu, Fengbin and Wang, Yiqi and Wu, Zihan and Wu, Weiwei and Liu, Leibo and Hu, Yang and Wei, Shaojun and Yin, Shouyi},
  booktitle={2023 IEEE International Solid-State Circuits Conference (ISSCC)},
  pages={254--256},
  year={2023},
  organization={IEEE}
}

@article{dao2022flashattention,
  title={Flashattention: Fast and memory-efficient exact attention with io-awareness},
  author={Dao, Tri and Fu, Dan and Ermon, Stefano and Rudra, Atri and R{\'e}, Christopher},
  journal={Advances in neural information processing systems},
  volume={35},
  pages={16344--16359},
  year={2022}
}

@article{vaswani2017attention,
  title={Attention is all you need},
  author={Vaswani, Ashish and Shazeer, Noam and Parmar, Niki and Uszkoreit, Jakob and Jones, Llion and Gomez, Aidan N and Kaiser, {\L}ukasz and Polosukhin, Illia},
  journal={Advances in neural information processing systems},
  volume={30},
  year={2017}
}

@article{dubey2024llama,
  title={The llama 3 herd of models},
  author={Dubey, Abhimanyu and Jauhri, Abhinav and Pandey, Abhinav and Kadian, Abhishek and Al-Dahle, Ahmad and Letman, Aiesha and Mathur, Akhil and Schelten, Alan and Yang, Amy and Fan, Angela and others},
  journal={arXiv e-prints},
  pages={arXiv--2407},
  year={2024}
}

@inproceedings{leviathan2023fast,
  title={Fast inference from transformers via speculative decoding},
  author={Leviathan, Yaniv and Kalman, Matan and Matias, Yossi},
  booktitle={International Conference on Machine Learning},
  pages={19274--19286},
  year={2023},
  organization={PMLR}
}

@article{liu2024deepseek,
  title={Deepseek-v3 technical report},
  author={Liu, Aixin and Feng, Bei and Xue, Bing and Wang, Bingxuan and Wu, Bochao and Lu, Chengda and Zhao, Chenggang and Deng, Chengqi and Zhang, Chenyu and Ruan, Chong and others},
  journal={arXiv preprint arXiv:2412.19437},
  year={2024}
}

@inproceedings{chen2023autodcim,
  title={Autodcim: An automated digital cim compiler},
  author={Chen, Jia and Tu, Fengbin and Shao, Kunming and Tian, Fengshi and Huo, Xiao and Tsui, Chi-Ying and Cheng, Kwang-Ting},
  booktitle={2023 60th ACM/IEEE Design Automation Conference (DAC)},
  pages={1--6},
  year={2023},
  organization={IEEE}
}

@inproceedings{wu2024next,
  title={Next-gpt: Any-to-any multimodal llm},
  author={Wu, Shengqiong and Fei, Hao and Qu, Leigang and Ji, Wei and Chua, Tat-Seng},
  booktitle={Forty-first International Conference on Machine Learning},
  year={2024}
}

@article{norrie2021design,
  title={The design process for Google's training chips: TPUv2 and TPUv3},
  author={Norrie, Thomas and Patil, Nishant and Yoon, Doe Hyun and Kurian, George and Li, Sheng and Laudon, James and Young, Cliff and Jouppi, Norman and Patterson, David},
  journal={IEEE Micro},
  volume={41},
  number={2},
  pages={56--63},
  year={2021},
  publisher={IEEE}
}

@article{gloeckle2024better,
  title={Better \& faster large language models via multi-token prediction},
  author={Gloeckle, Fabian and Idrissi, Badr Youbi and Rozi{\`e}re, Baptiste and Lopez-Paz, David and Synnaeve, Gabriel},
  journal={arXiv preprint arXiv:2404.19737},
  year={2024}
}

@article{wang2025syscim,
  title={SysCIM: A Heterogeneous Chip Architecture for High-Efficiency CNN Training at Edge},
  author={Wang, Shuai and Li, Ziwei and Ma, Yuang and Kang, Yi},
  journal={IEEE Transactions on Very Large Scale Integration (VLSI) Systems},
  year={2025},
  publisher={IEEE}
}

@inproceedings{park2024tp,
  title={TP-DCIM: Transposable Digital SRAM CIM Architecture for Energy-Efficient and High Throughput Transformer Acceleration},
  author={Park, Junwoo and Lee, Kyeongho and Park, Jongsun},
  booktitle={Proceedings of the 43rd IEEE/ACM International Conference on Computer-Aided Design},
  pages={1--8},
  year={2024}
}

@inproceedings{peng2019dnn+,
  title={DNN+ NeuroSim: An end-to-end benchmarking framework for compute-in-memory accelerators with versatile device technologies},
  author={Peng, Xiaochen and Huang, Shanshi and Luo, Yandong and Sun, Xiaoyu and Yu, Shimeng},
  booktitle={2019 IEEE international electron devices meeting (IEDM)},
  pages={32--5},
  year={2019},
  organization={IEEE}
}

\end{document}